\documentclass[12pt]{aastex}
\usepackage{emulateapj5,graphicx,times}

\shorttitle{Crosstalk in Suprime-Cam}
\shortauthors{Yagi}

\begin{document}

\title{Crosstalk Analysis of Suprime-Cam FDCCDs Using Cosmic Rays in Dark Frames} 

\author{Masafumi Yagi}
\affil{Optical and Infrared Astronomy Division, 
National Astronomical Observatory of Japan, 
Mitaka, Tokyo, 181-8588, Japan}
\authoremail{YAGI.Masafumi@nao.ac.jp}

\begin{abstract}
We analyzed the crosstalks in the new full depleted CCDs 
in the Subaru Prime Focus Camera(Suprime-Cam).
The effect is evaluated quantitatively using cosmic rays in dark frames.
The crosstalk is well approximated by a linear correlation 
and the coefficient is $\sim$$10^{-4}$.
The coefficients are not significantly different among the 10 CCDs.
We also find that the crosstalk appears not only in the corresponding
pixels but also in the next pixel but one. 
No crosstalk is detected in Suprime-Cam among different CCDs.
Based on the analysis, the correction procedure 
for the crosstalk is presented, and 
the application to the data is demonstrated.
\end{abstract}

\keywords{Data Analysis and Techniques -- Astronomical Instrumentation}

\section{Introduction}

Multi-channel CCD often suffers from
a crosstalk phenomenon between readout channels.
There are several user documents describing the crosstalk%
\footnote{http://www.noao.edu/kpno/mosaic/manual/mosa\_2.html}%
\footnote{http://www.astronomy.ohio-state.edu/MDM/MDM4K/}%
\footnote{http://www.stsci.edu/hst/acs/performance/anomalies/zoo\_xtalk.html}%
\footnote{http://www.ast.cam.ac.uk/ioa/research/vdfs/docs/reports/sv/}
and some observatories prepare data to correct this crosstalk%
\footnote{http://www.noao.edu/noao/mosaic/calibs.html}.
\citet{Freyhammer2001} estimated the effect in DFOSC
and FORS1 at the ESO VLT. 
The studies for Advanced Camera for Surveys (ACS) in 
Hubble Space Telescope (HST) is also available
\citep{Giavalisco2004a,Giavalisco2004b,Suchkov2010,Suchkov2012}.

\begin{figure}
\includegraphics[scale=0.45,bb=0 0 740 410]{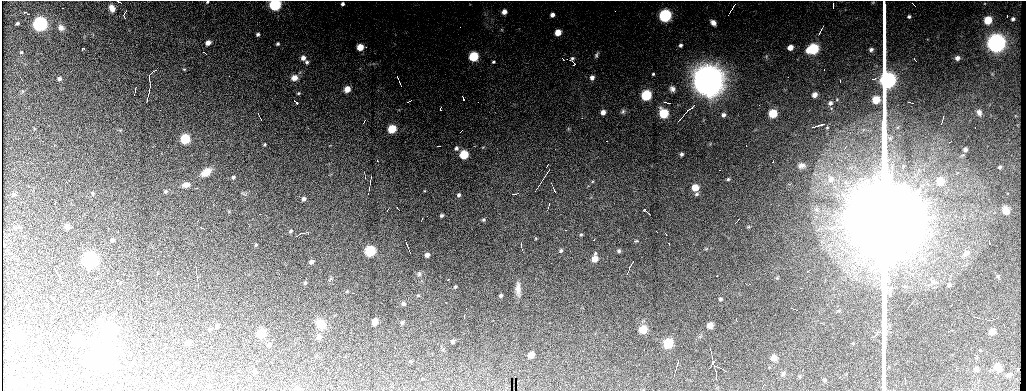}\\
\includegraphics[scale=0.45,bb=0 0 740 410]{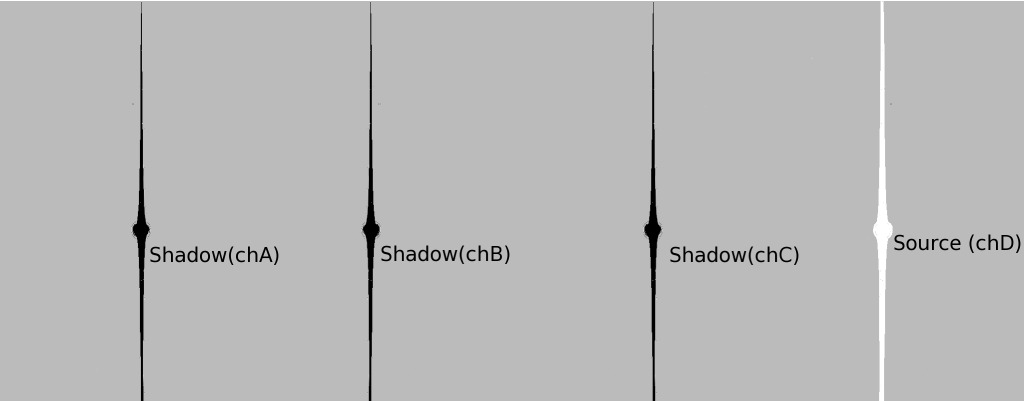}\\
\caption{An example of a crosstalk.
The bright star in chD makes shadows in chA, chB, and chC.
(top) The image is a 2048$\times$800 pixels cutout from 
a single exposure of a bias-subtracted and flat-fielded image.
(bottom) A schematic figure of the crosstalk source and 
its shadows of the top image.
}
\label{fig:crosstalk1}
\end{figure}

\begin{figure}
\includegraphics[scale=0.45,bb=0 0 803 168]{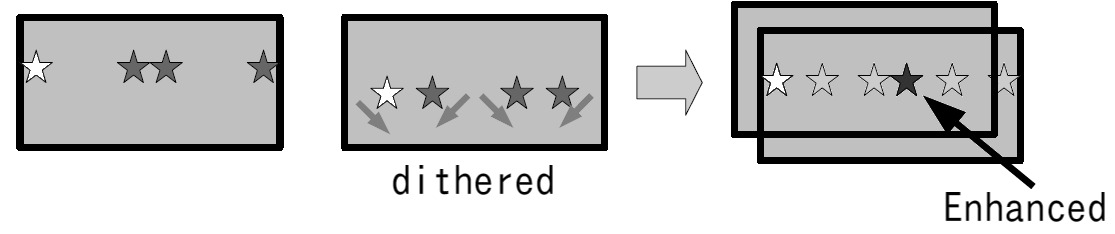}
\caption{A Schematic figures of the crosstalk effect after dithering.
(left) When a bright object is observed in chA,
shadows appears in chB, chC, and chD.
(middle) If the telescope pointing changed toward top-left,
the bright object moves toward bottom-right. 
Two of the shadows move toward bottom-left, while
one in the chC moves toward the bottom-right.
(right) As a result, the coadd of the two shots enhances 
the shadow in chC.
}
\label{fig:manga_crosstalk}
\end{figure}

Since the replacement of the CCDs in July 2008
with full depleted back illuminated CCDs 
\citep[FDCCD;][]{Kamata2008},
the data of Subaru Prime Focus Camera
\citep[Suprime-Cam;][]{Miyazaki2002} show crosstalk signatures in a CCD.
The effect is easily recognized
in narrowband data with low sky background
(Figure \ref{fig:crosstalk1}).
As one of the crosstalk 
dimming regions (shadows) 
has the same spatial parity as the source,
it obtains a higher signal-to-noise ratio(S/N) after coadding
(Figure \ref{fig:manga_crosstalk}),
and causes a problem even in a deep field study.

In this paper, we investigated the crosstalk effect in Suprime-Cam
in order to remedy this problem.
We adopted a method using cosmic rays in dark frames,
and present a recipe to remedy the crosstalk effects.

\section{Models and Method}

\subsection{Suprime-Cam}

\begin{figure}
\includegraphics[scale=0.3,bb=0 0 816 1171]{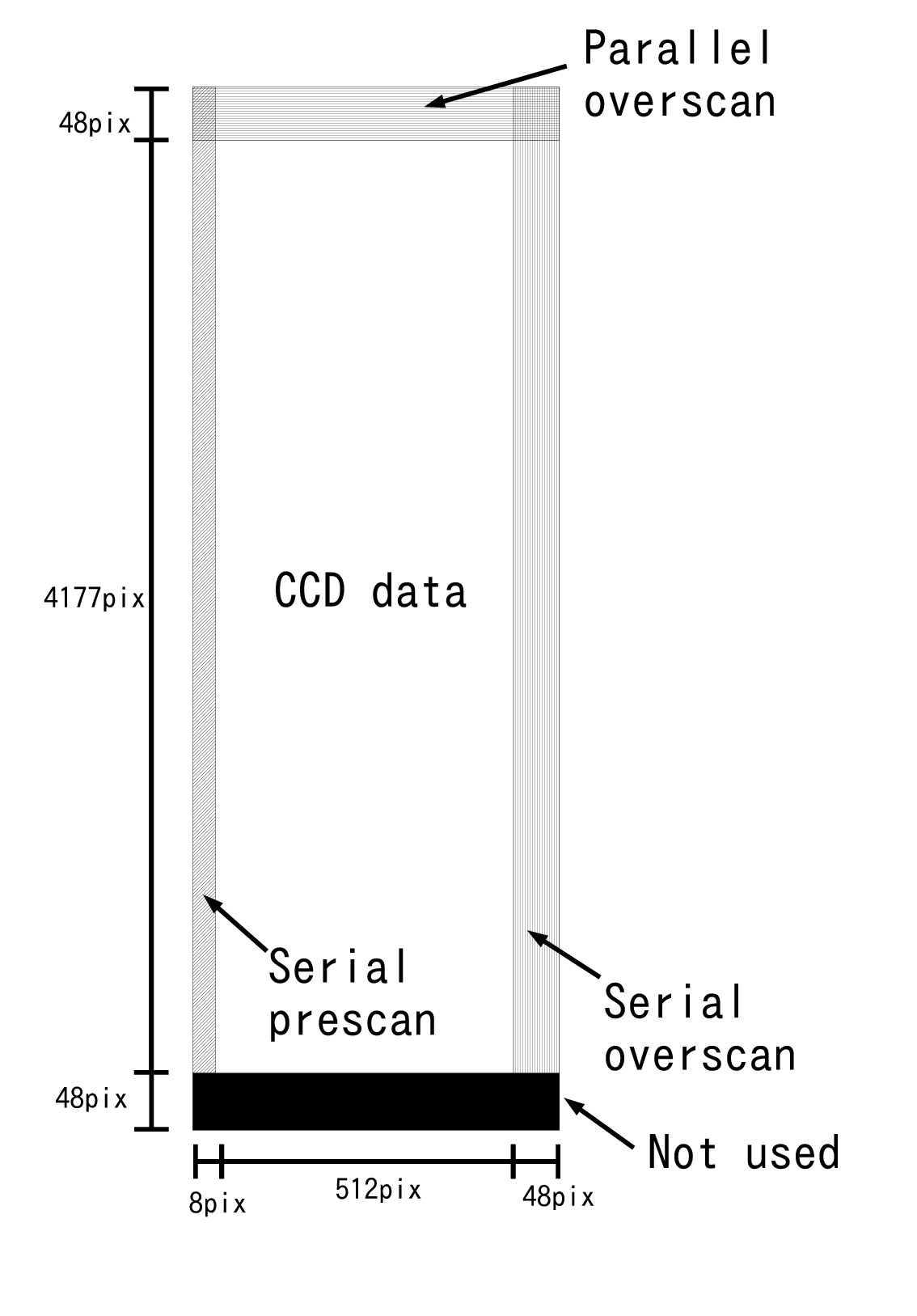}
\caption{
FITS data configuration of one channel of Suprime-Cam.
}
\label{fig:CCD1ch}
\end{figure}

Suprime-Cam is equipped with 10 FDCCDs,
and each FCCCD is read out from 4 channels.
In this paper, we call the channels chA, chB, chC, and chD
along the x-axis of the output FITS file for simplicity.
The data from each channel consist of 
512$\times$4177 CCD pixels,
8$\times$4177 pixels of the prescan region of serial read and 
48$\times$4177 pixels of the overscan region of serial read,
followed by (8+512+48)$\times$48 pixels overscan of parallel read.
The FITS data have an additional (8+512+48)$\times$48 pixels 
in the prescan region,
but it should not be used for the analysis (Figure \ref{fig:CCD1ch}).
The 10 CCDs are arranged in 2 rows of 5 CCDs each%
\footnote{http://www.naoj.org/Observing/Instruments/SCam/ccd.html}.
The CCDs in the upper row (detector ID=0,1,2,6,7) are
read from the top edge (y=4177), and those in the lower row 
(detector ID=3,4,5,8,9) are read from the bottom edge (y=1).

\begin{figure}
\includegraphics[scale=0.8,bb=0 0 330 220]{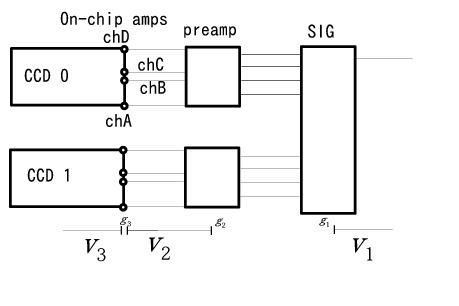}
\caption{A schematic figure of the readout system of Suprime-Cam,
and the names of signals used in this study}
\label{fig:manga_readout}
\end{figure}

When a CCD is read out, the charges stored in each pixel are 
converted to voltage at on-chip amplifiers(on-chip amps) at 4 channels.
The conversion factor of the on-chip amps has $\sim$15\% difference.
Suprime-Cam is equipped with 40 preamplifiers(preamps) arranged 
in 10 quad-channel preamp boards around the camera dewar.
One preamp board handles signals from a CCD.
The gain of the preamp is 3 \citep{Nakaya2008}
and the difference among the 40 preamps is $\sim$1\% 
(Nakaya, H. 2012; private communication).
The signals are then put into the signal board (SIG)
and a correlated double sampling is performed \citep{Nakaya2012}.
Suprime-Cam has 5 SIG boards, and one SIG board has 8 channels.
The signals from CCD0 and CCD1 are into the first SIG board,
those from CCD2 and CCD3 are into the next, and so on.
The signal is then converted into a digital value.
The analog-to-digital conversion factor 
is configured to be 1 \citep{Nakaya2008}, 
but slightly differs at each channel by $\sim$5\%.
The total gain of the Suprime-Cam data is the 
product of the three gains of the components;
the gain of the analog-to-digital converter (ADC) in the SIG
at each channel($g_1$),
the gain of the preamp ($g_2$), and 
the gain of the on-chip amp at each channel($g_3$).
A schematic figure is shown as Figure \ref{fig:manga_readout}.

The charges in CCDs are read out and converted to the digital data 
simultaneously in 40 channels. 
For example, when (x,y) of CCD0 is read, 
(1024-x,y), (1024+x,y), and (2048-x,y) of CCDs in the upper row
and (x,4178-y), (1024-x,4178-y), (1024+x,4178-y), and (2048-x,4178-y) 
of CCDs in the lower row are read at the same time.

\subsection{Crosstalk in Suprime-Cam}

The apparent crosstalk appears as follows.
When a bright object is observed at (x,y) in chA, 
shadows appear at three symmetric positions of saturated stars;
at (1024-x,y) in chB, at (1024+x,y) in chC, 
and at (2048-x,y) in chD, which are read out at the same time.
Currently, the crosstalk in the same CCD, the pixels read at the same time
show an apparent crosstalk. The possible crosstalk across the CCDs,
and effect on the adjacent pixels in the same CCD are examined later.

In the three shadows corresponding to (x,y),
(1024-x,y) and (2048-x,y), can be removed by dithering,
as their position at the sky changes
when the telescope pointing is changed.
However, the movement of the shadow at (1024+x,y) 
is the same as the object at (x,y),
and the shadow gains S/N when we coadd the dithered images 
(Figure \ref{fig:manga_crosstalk} right),
if the shadow is larger than the slight differential shift
by the optical distortion at the dithered pointing.
The existence of the shadow of the same parity
is different from quadrantic readout devices, 
such as ACS/HST and FORS1/VLT.
As investigated in a later section,
the shadow of the same parity causes problems 
in deep field imaging in broadband,
though the effect cannot be detected in a single exposure.

\subsection{Signal variables}

If the crosstalk occurs around the input of the on-chip amps, 
the effect would correlate with the raw charge
$v\times(g_1\times$$g_2\times$$g_3)$, where $v$ is 
the value in output FITS file after the bias subtraction.
If the crosstalk occurs between output of the on-chip amps
and input of the preamps, the effect would correlate with 
$v\times(g_1\times$$g_2)$, and if between output of preamps 
and input of ADC, the effect would correlate with 
$v\times$$g_1$. 

We estimated the relative value of the total gain 
$g_1\times$$g_2\times$$g_3$ and the
relative gain of SIG+preamps, $g_1\times$$g_2$,
and listed them in Tables \ref{tab:gain} and \ref{tab:SIGgain}.
The detail of the estimation is described in the Appendix.
Though we do not have $v_2$ nor $v_3$ data,
the difference of $g_1$ and $g_1\times$$g_2$ 
is small, since the difference of $g_2$ is only $\sim$1\% 
among the 40 preamps.

Using the relative gain values, 
we obtain three different signal values;
$v_1=v$, $v_2=v\times(g_1$$\times$$g_2)$, and 
$v_3=v\times(g_1$$\times$$g_2\times$$g_3)$.
$v_1$ is the count in FITS file in analog-to-digital unit(ADU),
$v_2$ is proportional to the voltage between the on-chip amps and 
the SIG, and $v_3$ is proportional to the charge in a pixel.

\section{Analysis}

\subsection{Method}
\label{sec:method}

We can evaluate the crosstalk by
taking the correlation between pixels in channel X and channel Y,
when the pixel in X has a large count and the pixel in Y has a smaller count.
A linear trend of crosstalk effect was reported by previous studies 
on other CCDs\citep[e.g.,][]{Freyhammer2001,Suchkov2010}.
Under the assumption that the crosstalk is linear to the source, 
the strength of the crosstalk is measured by the coefficient of 
proportionality.
The coefficient of other instruments are about $\sim$$10^{-4}$;
ACS/HST has -0.6..-2.3$\times10^{-4}$\citep{Suchkov2012}
and FORS1/VLT has -2.3..-2.5$\times10^{-4}$\citep{Freyhammer2001}.
It should be noted that the readout count suffers from 
the quantization error, and the difference smaller than 1 ADU 
in a certain pixel is buried in the noise.
Since the maximum output is $2^{16}-1=65535$ ADU in Suprime-Cam,
the crosstalk coefficient is meaningful when its absolute value is 
larger than $1/(2^{16}-1)\sim$$1.5\times$$10^{-5}$.

There are several types of data for measuring the correlation.
\citet{Freyhammer2001} used a calibration lamp with a mask on 
the focal plane and a night sky with standard stars.
\citet{Suchkov2010} used dark frames and sky frames.
In this work, we only use dark frames and cosmic rays
\footnote{We do not distinguish cosmic rays and hot pixels
and simply call them ``cosmic rays.''}
to minimize the error from flat fielding and the background level estimation.
The small spatial size of the high count pixels in cosmic rays, 
even smaller than the point spread function (PSF), 
enables us to examine a possible spatial extent of the crosstalk shadow over a pixel.
Moreover, the FDCCD of Suprime-Cam and the Subaru Telescope have 
several advantages for using cosmic rays in dark frames.
Thanks to the thickness of the FDCCD (250$\mu$m) 
and the high altitude of the Subaru Telescope, 
Suprime-Cam receives a relatively large number of cosmic rays.
The dark current is very low; less than 0.6 ADU per hour,
and the readout noise is also low; 2-2.5 ADU in rms
\citep{Kamata2008}. 
The acquisition of dark frames does not require either 
a special instrument setting such as a mask, nor a telescope time at night.
We can take a dark frames in a daytime if we can keep 
the instrument in the dark.
On the other hand, the drawback of this method is that 
it is not easy to obtain enough data of high-value count pixels.
For example, we could not investigate the behavior near
the full-well region in this study, because of the lack of such data.

\subsection{Data}

We used all dark frames with EXPTIME $>=$120 seconds
taken between 2008/12/03 and 2011/07/02 (UT); 
after the fix of the linearity problem, and 
before the hardware incident of the Subaru.
The data are retrieved from 
MASTARS\footnote{http://www.mastars.nao.ac.jp/}
and the 
SMOKA\footnote{http://smoka.nao.ac.jp/} archives.
The used frames are summarized in Table \ref{tab:darklist}.

\begin{figure}
\includegraphics[scale=0.4,bb=0 0 553 413]{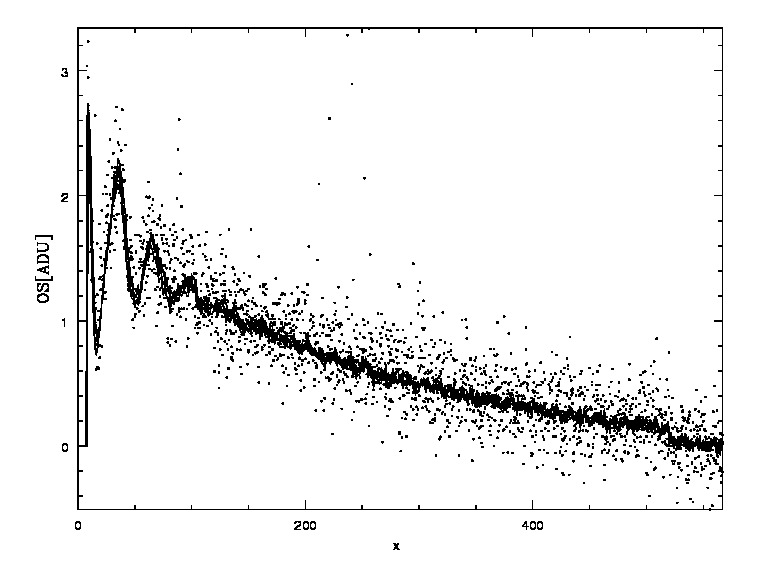}
\caption{Example of Y-overscan. 
The dots are the  parallel overscan of a frame(SUPA01181490)
after the serial overscan subtraction. 
The data in 4 channels are plotted.
The solid lines are the median of the parallel overscan function
of 4 channels of 10 CCDs of 10 exposures.
The prescan region (8 pixels at the left side) is discarded.
The overscan region is 48 pixels at the right side.
}
\label{fig:y-overscan}
\end{figure}

Bias is first subtracted using the median of 
48 pixels in the serial overscan region at each y.
The bias level has a $\sim$2 ADU waving pattern 
in x direction as shown in Figure \ref{fig:y-overscan}.
The pattern is common in all the CCDs, in all the channels,
and in all the exposures as far as we examined in the dark frames.
This pattern is corrected using the data in the parallel overscan
region. The median of the parallel overscan after 
the subtraction of the serial overscan of the parallel overscan 
reflects the pattern, and it is subtracted from the data in each frame.
Finally, the dark is subtracted,
and we call the value after the mean dark subtraction as $v_1$, hereafter.

The dark level is estimated by averaging the count in a channel 
avoiding the pixels which are hit by cosmic rays or 
are affected by the crosstalk because of a high count in other channels.
The mean dark count is various in different CCDs and channels.
The typical count of dark is (0.16$\pm$0.09) ADU for 300 second exposure
((5$\pm$3)$\times$$10^{-4}$ADU s$^{-1}$), and the rms is (2.2$\pm$0.2) ADU.
The rms includes readout noise and the error of bias/overscan subtraction.
As the total gain of Suprime-Cam is about 3--4 electron ADU$^{-1}$, 
the mean dark count is less than 1 electron.

In the bias and the dark corrected dark images, high count pixels are picked up
and then corresponding pixels in other 3 channels in the same CCD
are checked.
We arbitrarily adopted the threshold of the ``high'' count as 
300 ADUs in $v_1$. 
Total 1277705 pixels are marked as a high count pixel in the 370 frames.
The count of the three corresponding pixels to the high count pixel is 
cataloged.
When taking the correlation of $v_1(X)$ as the high count
and $v_1(Y)$ as the affected count, 
$v_1(Y)$ should be corrected the dark count, but $v_1(X)$ should not,
since the dark count would also contribute to the crosstalk.
However, as we take the $v_1(X)>300$ as the high count in following analysis,
the effect of the dark subtraction of 0.16 ADU makes 
0.05\% error in the coefficient. 
The error is negligible, as
we estimate the coefficients with 3 significant figures.
For simplicity, we used the bias and the dark subtracted value for $v_1(X)$
instead of the bias subtracted value.

\subsection{The fundamental variable}

\begin{figure}
\includegraphics[scale=0.4,bb=0 0 553 413]{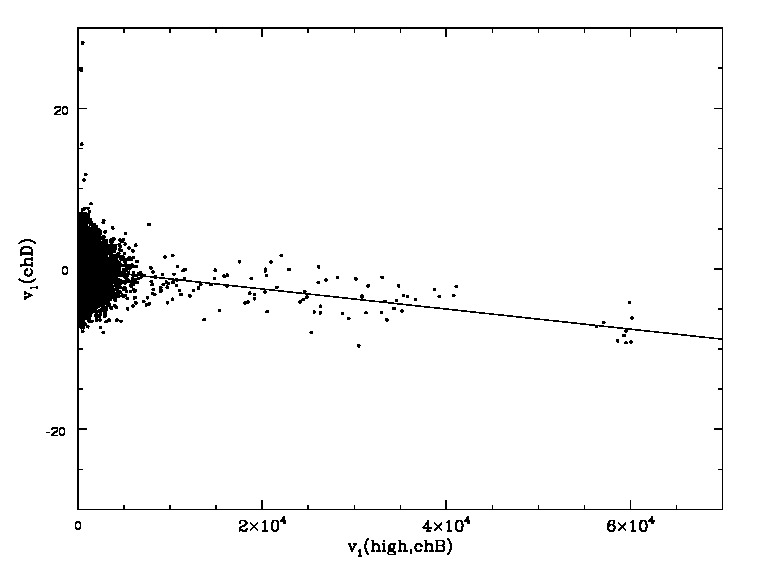}
\caption{
An example of crosstalk correlation.
The high count pixel value at chB of CCD0 versus
the pixel value at chD is plotted. All the exposures are used.
The best-fit regression to the data are
shown as the solid line.
}
\label{fig:ex1}
\end{figure}

In Fig. \ref{fig:ex1} an example of the correlation is shown.
The data are CCD0 and a high pixel is at chB and the checked the
corresponding pixel in chD. 
Apparent linear correlation is recognized.
The result shows that the shadow appears even in the 
dark data with negligible ($<1ADU$) background charges.
It suggests that the crosstalk phenomenon in Suprime-Cam 
should be a slight shift of the zero level.
This assumption is consistent with the result by 
\citet{Giavalisco2004a} that the effect is additive and not multiplicative.
We can therefore expect that an additive correction established 
with these negligible background data is also valid for 
the object images with sky backgrounds.

We then multiply two kinds of relative gain to $v_1$.
One is the value multiplied by the SIG+preamp gain in 
Table \ref{tab:SIGgain}. 
It represents the voltage between the output of the on-chip amp 
and the input of SIG. We call it $v_2$. 
The other is the value multiplied 
by the total gain in Table \ref{tab:gain}, 
which is proportional to the photo-charges. 
We call it $v_3$.
The question is which is the fundamental variable,
$v_1$, $v_2$ or $v_3$.
As the gains are different among channels, 
the behavior of the three variables is different.
A clue is $\sim$14\% change of gain of the on-chip amp of chA of CCD9 in 2010/10.
If crosstalk depends on photo-charges ($v_3$), 
the coefficient of the crosstalk should change if the gain of on-chip amp($g_3$)
changes. If crosstalk does not depend on the gain,
the coefficient should remain the same. 

We calculated the regression line using the data before 
the change of the gain, and tested whether 
the data after the change follow the same regression line.
For the regression, 
we estimated the distribution of $v$(other), especially 
the fraction of outliers. 
The distribution of the pixel values which are not affected 
by the high count pixels is well approximated by Gaussian in 
$-5\sigma<v<5\sigma$, with several ($\sim$$1.1\times10^{-4}$) outliers
on the positive side. 
The fraction is much larger than the expected fraction of $\leq$$5\sigma$ 
in Gaussian, $5.7\times10^{-7}$, and possibly due to 
weak radiation events.
If we exclude $\geq$$5\sigma$ data in normal distribution,
the expected reduced $\chi^2$ is only $1.5\times10^{-5}$ smaller.
Therefore, we neglect the effect of the truncation as
$\chi$$\leq 5\sigma$.

We restricted the data that input of high count pixel is in chD and 
output is in chA in order to exclude the effect of the difference 
of the coefficients of different combination of the channels
discussed in the next section.
We adopted chD because it has more high count pixels 
than chB or chC by chance in our data.
From likelihood-ratio test with the critical value of 5\%,
we obtained the result that the regression of $v_1$ and $v_2$ 
is not significantly different, while the regression of $v_3$
changes significantly after the change of the gain.

\begin{figure}
\includegraphics[scale=0.4,bb=0 0 412 700]{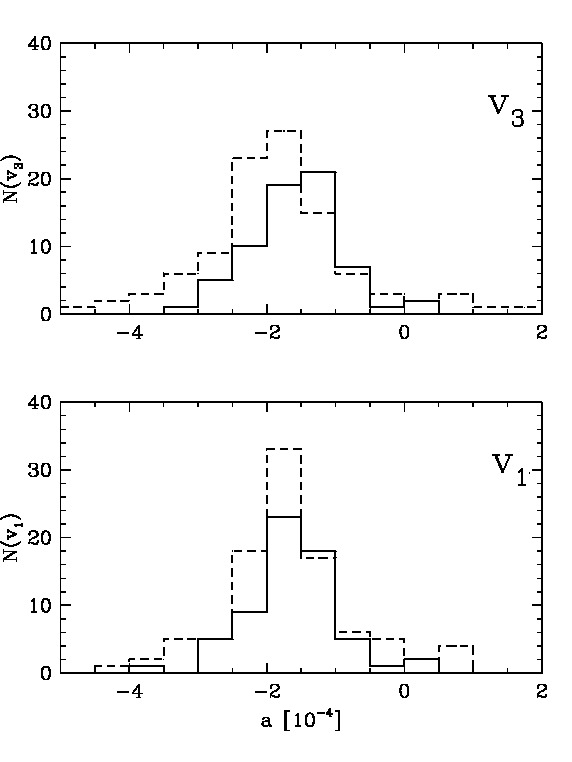}
\caption{
Histogram of $a=v$(other)/$v$(high) of CCD9
for high count pixel is chD, output is chA and $v$(high)$>$15000 data.
The solid line is before the gain change and the broken one is
after the change. The top panel is $v_3$ and the bottom panel is $v_1$.
The shape of the histogram of $v_3$ varied with the change of the gain,
while that of $v_1$ was invariant.
}
\label{fig:chip9hist}
\end{figure}

As another check, we plotted the histogram of  
$a$=$v$(other;chD)/$v$(high;chA) 
in $v$(high;chA)$>$15000 data of $v_1$ and $v_3$
in Fig. \ref{fig:chip9hist}.
We can see a shift of the histogram of $v_3$, while 
$v_1$ remains the same.
We therefore conclude that the crosstalk affects 
$v_1$ or $v_2$, and not $v_3$.
The result resembles the result by \citet{Freyhammer2001},
who noted that
``Changing the gain, e.g. from low to high gain, does not alter 
 the cross-talk amplitudes, when the cross-talk originates from 
 the CCD itself rather than from the ADCs electronic circuits.''

The result implies that the crosstalk does not occur
inside of the CCD but downstream of output of the on-chip amp.
In the following analyses, we do not use $v_3$.
Whether $v_2$ is more fundamental than $v_1$ or not 
cannot be distinguished by this CCD9 gain analysis.
As the difference of $v_1$ and $v_2$ is 
7\% at most as listed in Table \ref{tab:SIGgain},
the correction of the crosstalk may have 
a comparable error if the wrong variable is used.

\subsection{Variation of crosstalk coefficients}

\begin{figure}
\includegraphics[scale=0.4,bb=0 0 553 413]{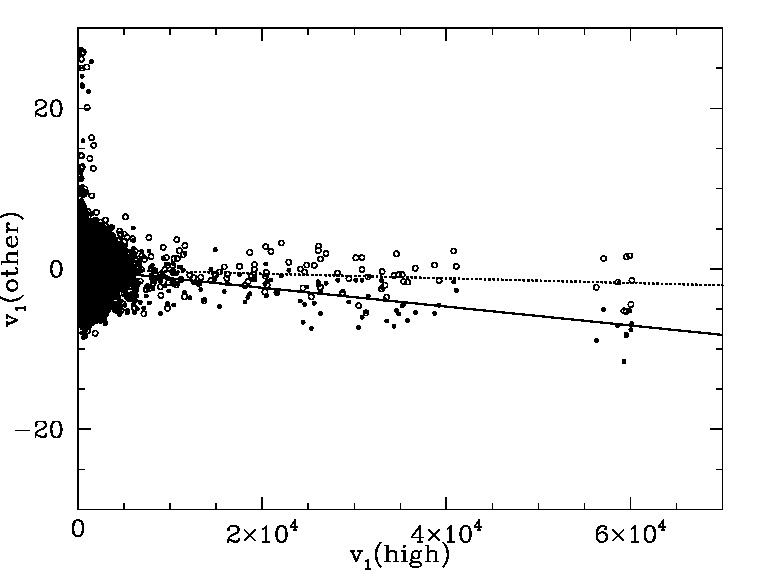}
\caption{
The high count pixel at $v_1$ of chB vs 
$v_1$ of chA (filled circles)
and chC (open circles) of CCD0.
The best-fit regression to the data are
shown as the solid line(chA) and the dotted line(chC).
}
\label{fig:chBchCplot}
\end{figure}

\begin{figure}
\includegraphics[scale=0.5,bb=0 0 490 324]{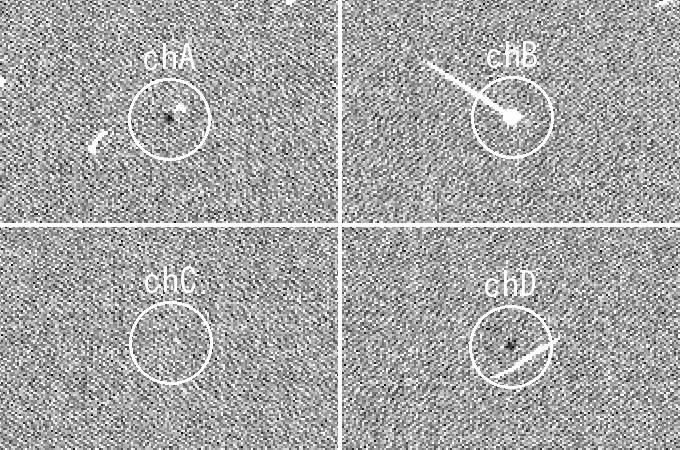}
\caption{
An example of different coefficients among channels.
The images are cutouts of a dark frame.
The color scale is the same for the four.
A cosmic ray hit in chB, and the shadow appears in
chA and chD. 
In chC, a shadow is not recognized 
around the corresponding pixel.
}
\label{fig:chBchCexample}
\end{figure}

We investigated the linear correlation 
of each CCD of each combination of the channels,
because we noticed that some combination of channels have
weaker crosstalk signal than others.
For example, the chC affected from chB in CCD0 
shows weak crosstalk signal as shown in Fig.
\ref{fig:chBchCplot} and Fig. \ref{fig:chBchCexample}.

Ideally, we can estimate all combinations 
separately by investigating all possible datasets.
In our data, however, some combinations do not have 
enough data at large $v$(high) and the error of 
the estimation of the coefficient is large.
We therefore set a working hypothesis and test its validity.
The hypothesis is that the coefficient is the same 
for the same combination in a mirror symmetry.
For simplicity, we will call the data where 
a high count pixel is in chX and the output is chY as crosstalk of XY,
and write it as cXY. 
From the assumption of the mirror symmetry, 
the combinations are reduced to 4 groups: 
4 of cAB-like combinations, which include cAB, cBA, cCD, cDC, 
4 of cAC-like ones, 2 of cAD-like ones, and 2 of cBC-like ones. 
We call the groups as gAB, gAC, gAD, and gBC, respectively.
As the small $v$(high) data only add the noise, 
we restricted the fitting range at $v$(high)$>5000$.
We also tested $v$(high)$>15000$ but the difference between them is small.
The coefficients of the best-fit regressions are 
shown in Table \ref{tab:coeff1} and \ref{tab:coeff2}.
The 95\% confidence intervals are calculated from likelihood-ratio test.
The regression of gAB and gAC have an overlap of
the confidence intervals, while other two do not have the overlap.

Then, data of each channel pair are compared with the best-fit function
by likelihood-ratio test with the critical value of 5\%.
One cBC data (CCD5) and two cBC data (CCD0 and CCD5) are 
significantly different from the best-fit function 
both in $v_1$ and $v_2$.
As the total number of combinations are 120, the expected number 
of significantly different pairs should follow binomial distribution
Bi(120,0.05) in an ideal case.
We therefore conclude that the coefficients of the crosstalk
are not significantly different in the same symmetry group,
and we mix all the chips and the combinations 
in the same symmetry group hereafter.
In future, this hypothesis should be re-examined 
when more data are available.

As the coefficient of gBC in the same CCD is different 
from other three, we can assume that the crosstalk may 
occur around the output of on-chip amp, because
the difference of gBC from other three groups
exists only inside the CCD package.
This implies that $v_2$ would be the fundamental variable.
We therefore use $v_2$ hereafter.

\subsection{Profile along x-axis}

We checked whether the crosstalk occurs only in the 
corresponding pixel to the high count pixel.
If the crosstalk has a time duration,
not only the pixels which were read at the same time but
also the pixels which were read later might be affected.
For example, if (x,y) in chA has a high count,
(1024-(x+$\Delta$x),y) is examined in chB,
and also the corresponding pixels in chC and chD.
We should be careful in the analysis that
if ((x+$\Delta$x),y) in chA is also a high count pixel, 
the shadow at (1024-(x+$\Delta$x),y)
would be caused by a normal crosstalk effect from ((x+$\Delta$x),y).
We therefore set an additional constraint that 
((x+$\Delta$x),y) and corresponding pixels in other three channels 
should not be high count pixels, i.e. $v<$300.
We then found a sign that the confidence intervals of the coefficient
are significantly different from 0 at $\Delta$x$>0$.

\begin{figure}
\includegraphics[scale=0.3,bb=0 0 900 600]{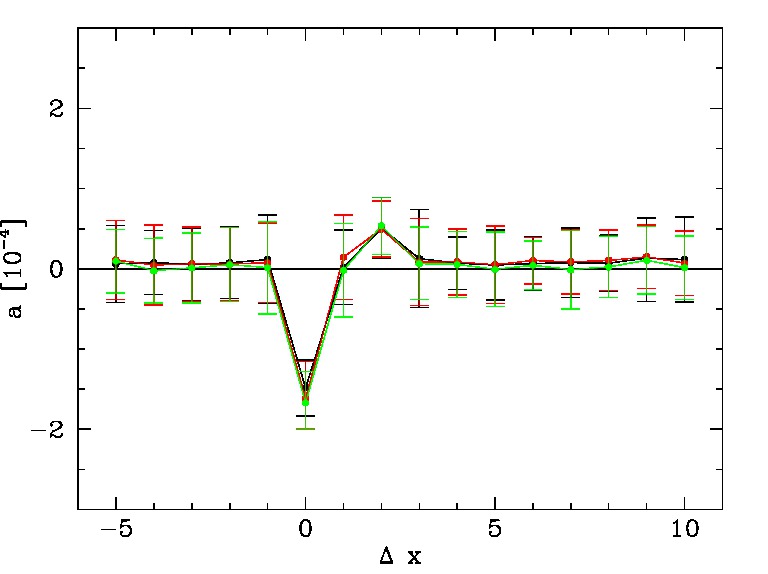}
\includegraphics[scale=0.3,bb=0 0 900 600]{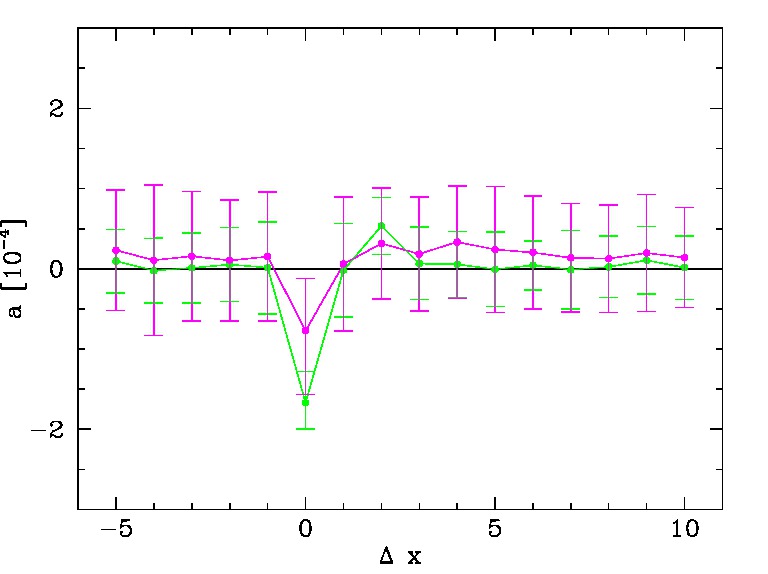}
\caption{
The crosstalk coefficients of adjacent pixels in $v_2$.
Only the high count pixels which extend 1 pixel along the x-axis are used,
and therefore the error is larger than the values in Table \ref{tab:coeff1}.
The errorbars show 95\% confidence intervals.
At x=0, the high count pixel is read.
The data read before the the high count pixel is read are
shown in x$<$0, and those after the high count pixel is read are in x$>$0.
(left) The profile of gAB, gAC, and gAD are plotted 
in black, red, and green, respectively.
The coefficient is negative at x=0, which corresponds to
the shadow. The three profiles are similar.
(right) The profiles of gAD and gBC are plotted 
in green and magenta.
The coefficient of gBC is nearly zero.
}
\label{fig:xprof_n1}
\end{figure}

For a detailed study, 
we select the high count pixels which extend only one pixel
along the x-axis, and calculate the profile 
of the crosstalk coefficients along the x-axis.
The result is shown in Figure \ref{fig:xprof_n1},
and the coefficients around x=0 are given in Table \ref{tab:xprof_n1}.
The errorbars represent the 95\% confidence intervals.
The readout sequence is toward +x.
The pixel at $\Delta$x=1 is read out just after the high count pixel is
read out, for example.
The profile of gAB, gAC, and gAD resemble one another.
In $\Delta$x$<0$ region, the coefficient is $\sim$ 0.
At $\Delta$x=0, the coefficient is significantly negative.
It is the shadow of the crosstalk in Figure \ref{fig:crosstalk1}.
Then the coefficient is back to almost 0 at $x=1$, 
and then {\it significantly} positive at $x=2$.
It resembles a damped oscillation pattern.
On the other hand, the profile of gBC shows
no significant crosstalk except at x=0.

\subsection{Crosstalk across the CCDs}

If the crosstalk occurs around SIG, the crosstalk of 
CCD(2n) and CCD(2n+1) may occur,
as they are handled in the same SIG board.
The possibility is examined in the same way as in the previous section.
We picked up a high count pixel and check the corresponding 
pixels which are read at the same time in other CCDs.
None of the coefficients of the crosstalk across the CCDs is
significantly different from 0. 
If we use all the combinations, the coefficient is 
$-0.01\times$$10^{-4}$, and their 95\% confidence interval is 
$-0.11\times$$10^{-4}<a<0.08\times$$10^{-4}$.
The result supports the assumption 
that the crosstalk would occur around on-chip amp.
And the possible small crosstalk across the CCDs is negligible, 
because the crosstalk effect is buried in the quantization noise 
if the coefficient is smaller than $\sim$$1.5\times$$10^{-5}$
as discussed in section \ref{sec:method}.

\section{Correction of the crosstalk}

\subsection{Procedure}
We tried to correct the two significant pixels at $x=0$ and $x=2$
using the coefficients in Table \ref{tab:xprof_n1}.
The recipe is as follows:
\begin{enumerate}
\item Prepare an overscan subtracted image.
\item Convert the pixel value ($v_1$) to $v_2$ using the 
 gain in Table \ref{tab:SIGgain}. 
\item Visit each pixel in the frame.
\item When visiting (x,y) in chA, 
   calculate the crosstalk effect from (1024-x,y) as $\Delta$ x=0
   and (1024-(x+2),y) as  $\Delta$ x=+2 in chB.
   The crosstalk from chC and chD are also calculated.
\item Subtract the sum of the effect from the pixel value at (x,y)
   in the output frame.
\end{enumerate}

This procedure is based on several assumptions.
First, we assumed that the effects from different pixels are additive.
The coefficients we use are calculated from isolated signals.
We assumed that the crosstalk effect is a simple sum
when several high count pixels are connected in the x-axis.
This assumption is checked later.
Second, we assumed that the crosstalk from 
different channels is also additive.
This must be verified by checking whether the crosstalk signal
changes when two or more pixels have high count.
For example, we should check whether the crosstalk in chA is 
doubled or not if chC and chD are hit by cosmic rays.
In current data, such events are too few to make a conclusion. 
Third, we assume that the crosstalk does not affect 
the pixels in the same channel. 
It is difficult to check, since the local change of bias level 
from $\Delta$x=0 cannot be distinguished from a change of the gain and the
effect would be very small ($\sim$$10^{-4}$).
The $\Delta$x=+2 signal might be seen in the same channel,
but we cannot distinguish the crosstalk signal from
the original signal, as the intrinsic profile 
of the cosmic rays is unknown.

\begin{figure}
\includegraphics[scale=0.3,bb=0 0 900 600]{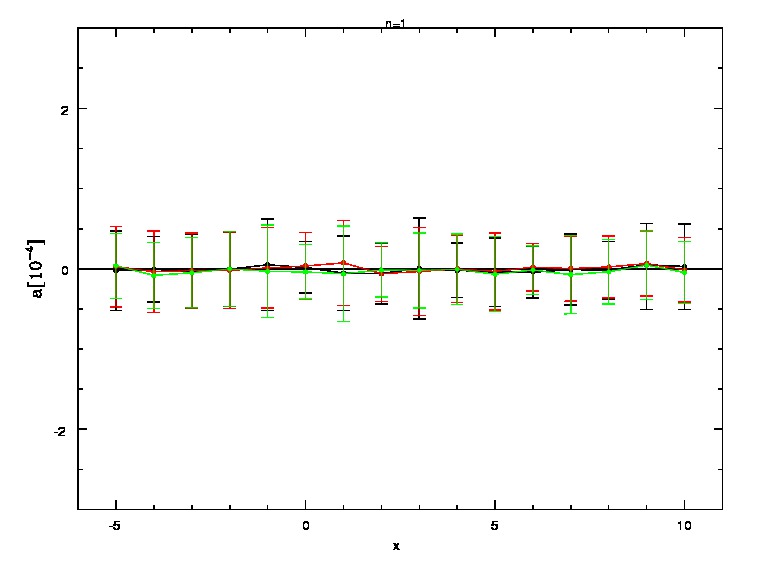}~
\includegraphics[scale=0.3,bb=0 0 900 600]{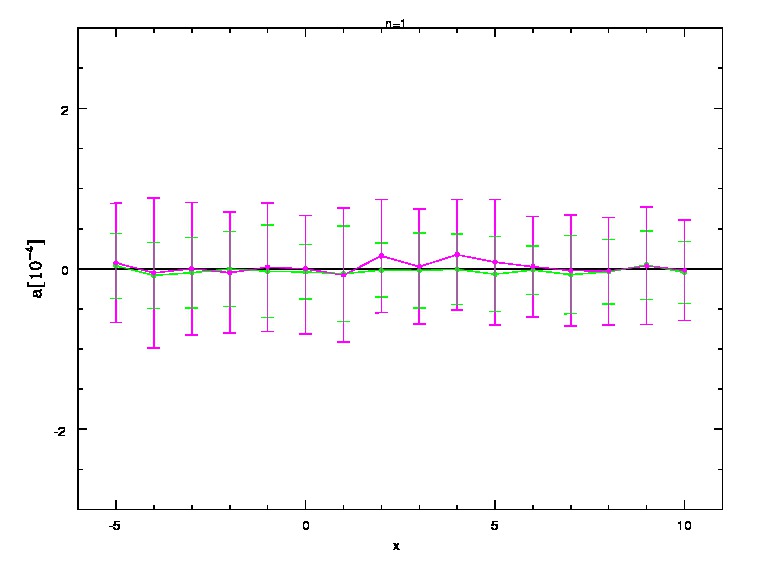}\\
\includegraphics[scale=0.3,bb=0 0 900 600]{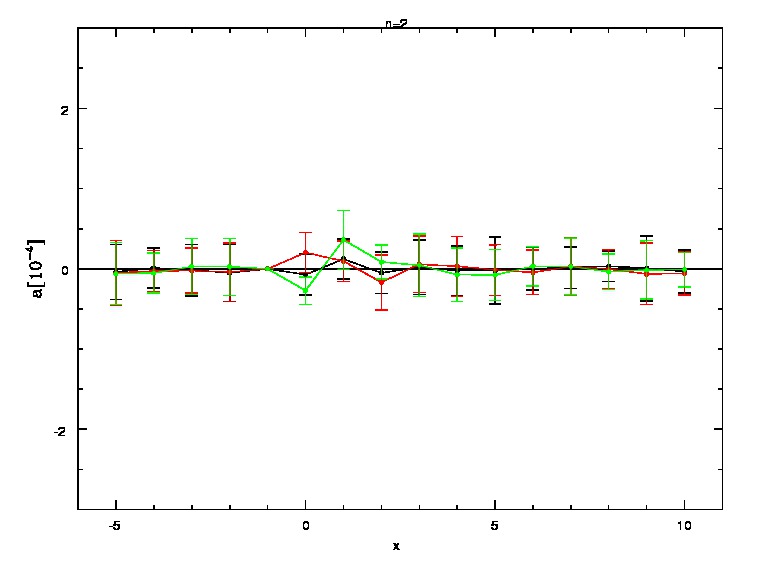}~
\includegraphics[scale=0.3,bb=0 0 900 600]{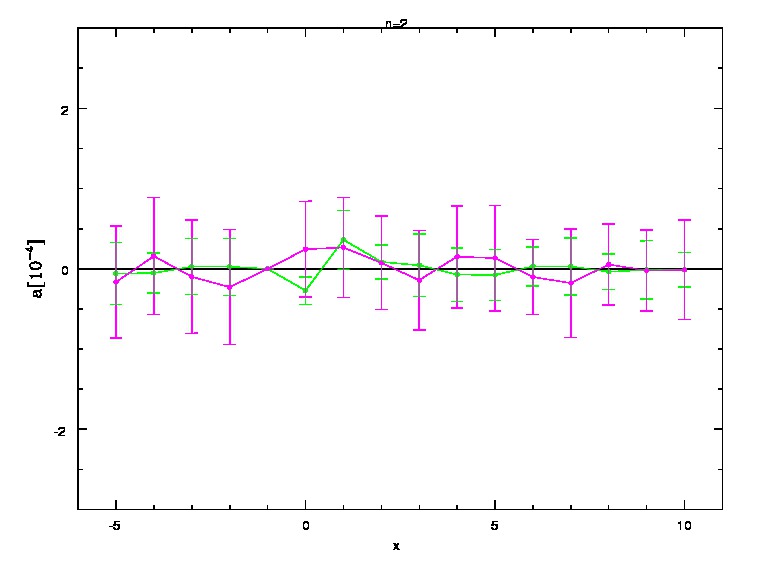}\\
\includegraphics[scale=0.3,bb=0 0 900 600]{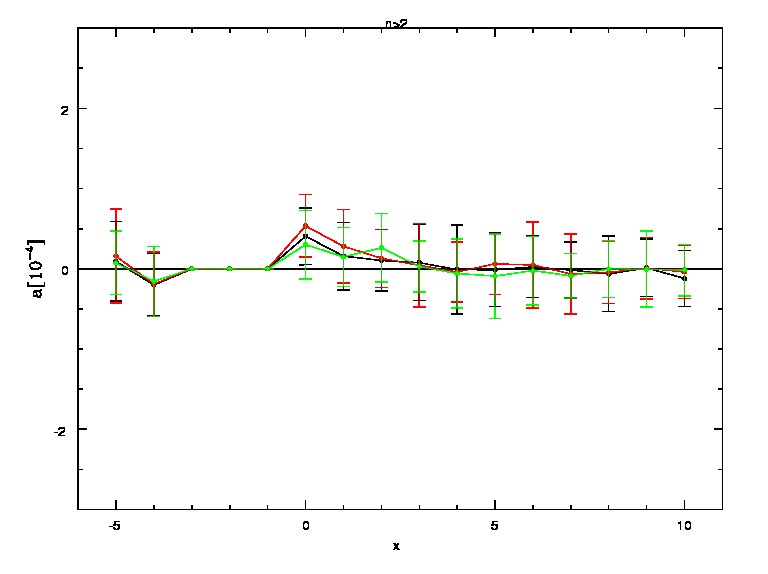}~
\includegraphics[scale=0.3,bb=0 0 900 600]{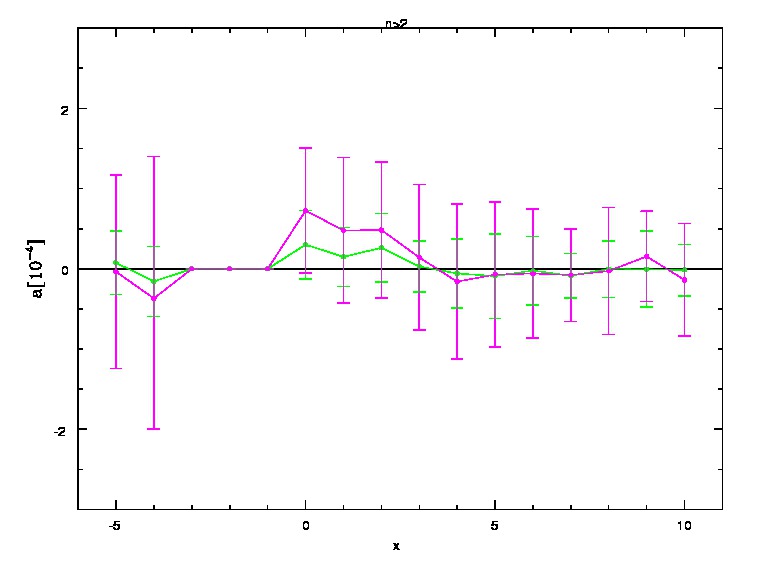}\\
\caption{
The crosstalk coefficients of adjacent pixels in $v_2$
after our correction is applied.
Top) The coefficients using high count pixels which extend 1
pixel along the x-axis.
The color allocation is the same as Fig. \ref{fig:xprof_n1}. 
Middle) The coefficients using high count pixels which extend 2 pixels.
x=0 is the last one of the connected pixels. As x=-1 is contaminated 
by the high count pixel, there is no data and coefficient is set to 0.
Bottom) The coefficients using high count pixels which extend
more than 2 pixels. The coefficients at x=-1, and x=-2 is 0
because of no data.
}
\label{fig:xprof_after}
\end{figure}

We checked the first assumption,
the effects from different pixels are additive,
using the crosstalk corrected images following the recipe.
The profile after the correction is shown in Fig \ref{fig:xprof_after}.
As expected, the crosstalk is corrected well for single high count pixels,
and most of the coefficients are consistent with no crosstalk ($a=0$).
However, connected high count pixels show a significant sign that
pixels at x=0 in some combination have a significant signal.
This means that the crosstalk phenomenon is not perfectly additive.

Recently, \citet{Nakaya2012} reported that preamp+SIG 
have a remnant signal of $\sim$1 ADU 
in following pixels in 4 channels after 50000 ADU signal in a channel.
It corresponds to $\sim$$2\times$$10^{-5}$ of the coefficient
at $x=1$ in our analysis.
The different behavior of crosstalk after connected 
high value pixels may be a combined effect of the crosstalk and 
the remnant signal.
Currently, it is difficult to investigate further 
because of the lack of sufficient data.
Detailed investigation on this point will be possible when
more data are available in future.

\subsection{Application to the data}

In the previous section, we get a recipe for the 
crosstalk correction for data with a negligible background.
In this section, we will test the correction
to the data with a background.
We will apply the correction to two kinds of astronomical data.
One is a narrowband image with a bright star.
Such data has a low background sky level, and 
the effect of the crosstalk is apparent in a single image.
The other is a deep field taken with a broadband filter.
It is difficult to recognize the crosstalk in a single image,
but the coadd enhances the S/N of the shadow.
Then, we will estimate the effect of the crosstalk 
correction quantitatively.

\subsubsection{Narrowband images}

\begin{figure}
\includegraphics[scale=0.6,bb=0 0 459 253]{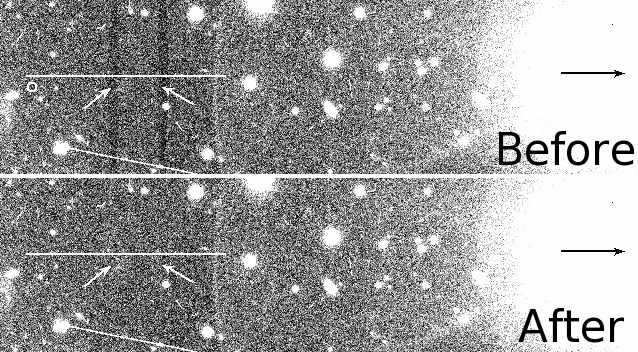}
\caption{The effect of the correction for a narrowband image.
The top panel is a cutout of a flat-fielded image and 
the bottom panel is that of a crosstalk-corrected and flat-fielded image.
The black arrow shows a position of the saturated blooming in chD, 
and the white arrows show corresponding shadows in chB and chC 
by the crosstalk.
The horizontal line shows the position of the profile 
shown in Fig \ref{fig:M83xprof},
and the circle in the top panel indicates the 2 arcsec aperture size.
}
\label{fig:M83}
\end{figure}

\begin{figure}
\includegraphics[scale=0.6,bb=0 0 553 413]{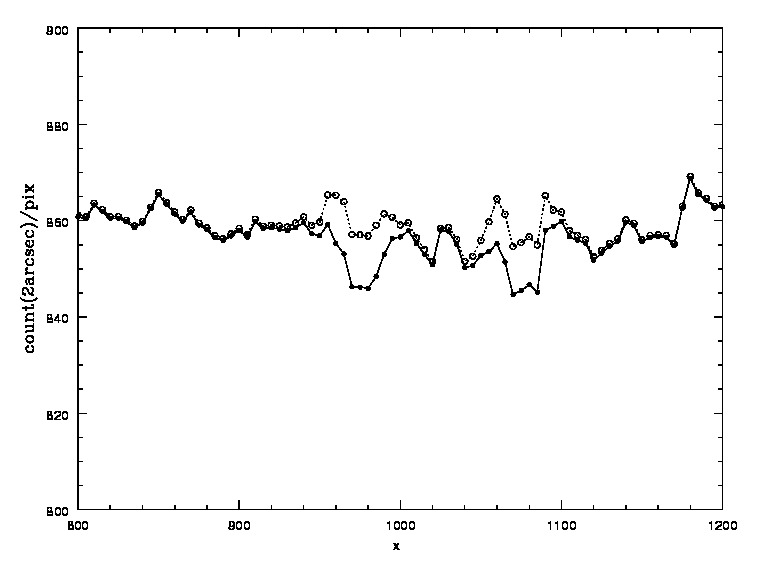}
\caption{The surface brightness profiles of 
in 2 arcsec apertures along the x-axis.
The images are Fig \ref{fig:M83}.
The solid line with filled circles is the profile
before the correction and the broken line with open circles
is that after the correction.
Around x=980 and x=1080, the shadow is corrected.
}
\label{fig:M83xprof}
\end{figure}

For the first test, 
we used H$\alpha$(N-A-L659) data of M83 
of 720 seconds exposures, which were used in \citet{Koda2012}.
After flat-fielding, the background is typically 700 -- 900 ADU, 
and rms is 20 -- 40 ADU in a pixel.
Saturated stars make blooming of $\sim$60000 ADU, 
and the shadow will be $\sim$10 ADU.
An example is shown in the top pane of Fig \ref{fig:M83}.
Though the crosstalk signal is 0.2 -- 0.5$\sigma$ in a pixel,
the corresponding shadow pattern is recognized in a single image,
because the blooming pattern has a width of $\sim$20 pixels.
The result of the correction is shown in the bottom pane of Fig \ref{fig:M83}.
The apparent shadow is corrected by our recipe.
Fig \ref{fig:M83xprof} is a surface brightness profile in 
2 arcsec apertures along x-axis of the two images.
The shadow is corrected well.

\subsubsection{Deep field}

For the second test,
we adopted a part of z-band(W-S-Z+) data of UV4a field.
The exposure time is various between 180 sec to 720 sec.
The median of the sky level is $\sim$26000 ADU.
The position angle is the same for all the exposures,
and the crosstalk shadows of the same parity overlap at the same position.
We performed a standard reduction of Suprime-Cam.
After the reduction, the shadow is not obvious in a single exposure,
because of the high photon noise.
However, it becomes apparent after the coadd of many exposures.

\begin{figure}
\includegraphics[scale=0.5,bb=0 0 323 313]{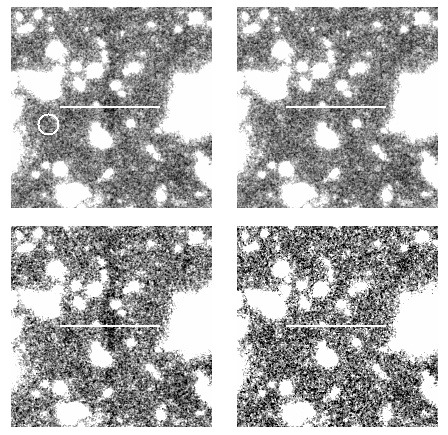}
\caption{
An example of the crosstalk effect in deep field data 
taken in broadband.
The color scale is the same for the four panels.
The horizontal line shows the position of the profile 
shown in Fig \ref{fig:UV4a_prof},
and the circle in the top-left panel indicates the 2 arcsec aperture size.
(top left) A cutout of a clipped mean coadd of the images without the
crosstalk correction. A vertically elongated shadow is recognized.
(top right) The same as the top-left but with the crosstalk correction.
(bottom left) A clipped mean coadd of a subsample.
The frames which would make the shadow at this position are used.
(bottm right) The same as the bottom-left but those
that would not create a shadow at this position are used.
}
\label{fig:UV4a}
\end{figure}

\begin{figure}
\includegraphics[scale=0.5,bb=0 0 553 413]{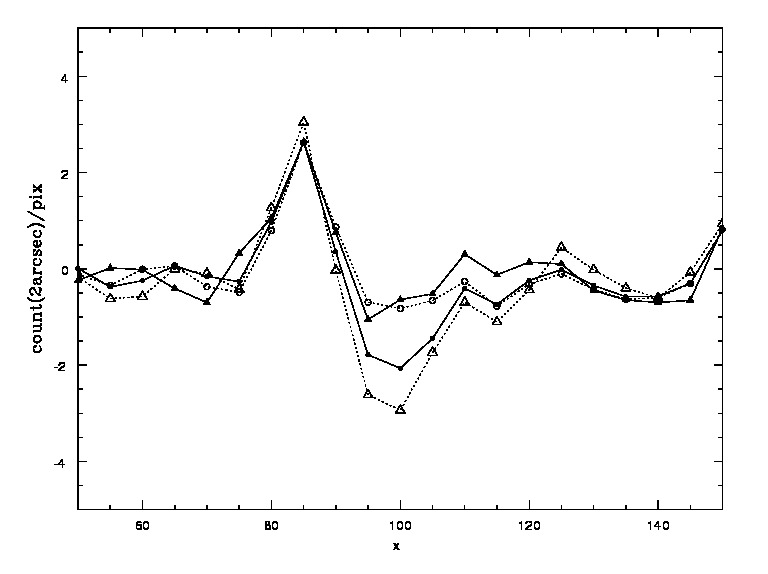}
\caption{
The surface brightness profiles of 2 arcsec apertures along the x-axis
shown in Fig \ref{fig:UV4a}.
The solid line with filled circles is before the correction,
and the broken line  with open circles is after the correction.
The broken line with open triangles and 
the solid line with filled triangles 
represent the coadd of affected frames and that of unaffected frames
respectively.
}
\label{fig:UV4a_prof}
\end{figure}

We picked up an example of the shadow 
as shown in Figure \ref{fig:UV4a} top-left pane,
and the coadd of the corrected data with our recipe is 
the top-right pane.
Then, we divided the data into two subgroups according to
the x-position of the dithering so that the celestial position 
shown in Figure \ref{fig:UV4a} is affected by the crosstalk 
in the frames of one of the groups, and not in the frames of 
the other group.
The numbers of the frames are 68(affected) and 86(not affected).
The result of the coadd of each group is also shown in 
Figure \ref{fig:UV4a} bottom-left and bottom-right.
The surface brightness profile of 2 arcsec apertures along the x-axis 
is shown as Figure \ref{fig:UV4a_prof}.
It shows that the shadow at x=100 is corrected well by our method 
(open circles), and the spatial profile of the coadd of 
the non-affected frames (filled triangles) is recovered.

\section{Summary}

Using cosmic rays in dark frames,
we evaluated the crosstalk in the new Suprime-Cam FDCCDs.
The strength of the crosstalk is not affected by 
a change of the GAIN of the on-chip amps,
which implies that the crosstalk occurs not 
inside the CCD but downstream from the output of the on-chip amp.
The crosstalk effect is well approximated by a linear correlation. 
The coefficient seems to be correlated with the 
distance between the on-chip amplifiers in the CCD,
which implies that the crosstalk occurs around on-chip amps.
The coefficients are not significantly different among the 10 CCD CCDs.
No crosstalk is detected among the different CCDs.
We also find that the crosstalk appears not only the corresponding
pixels but also at the next pixel but one. 
We present a recipe to remedy the crosstalk effect.
The recipe is applied to the real data to show that it works well.

\acknowledgments

We thank the anonymous referee for suggestions which 
helped us to improve and clarify the manuscript.
We appreciate Fumiaki Nakata for obtaining dark frames,
Hidehiko Nakaya for the detailed information
on readout system of Suprime-Cam, 
Yukiko Kamata and Hisanori Furusawa for the previous 
GAIN measurement of the CCDs.
We appreciate Yutaka Komiyama and Satoshi Kawanomoto 
for suggestions and comments on the analysis.
This work is based on archived data collected at the Subaru Telescope.
The data are obtained from MASTARS, which is operated by the Subaru Telescope, 
and the SMOKA, which is operated by the Astronomy Data Center,
National Astronomical Observatory of Japan.

\begin{appendix}

\section{Relative gain estimation}

Data of Suprime-Cam has GAIN information of each channel in FITS header.
The values should be $g_1\times$$g_2\times$$g_3$,
where $g_1$, $g_2$, and $g_3$ are
the gain of the analog-to-digital converter (ADC) in the SIG
at each channel, 
the gain of the preamp of each CCD,
and the gain of the on-chip amp at each channel,
respectively.

The header values, however, are known to have large error
\footnote{http://smoka.nao.ac.jp/help/help\_SUPnewCCD.jsp}.
The reason was that the preamp and the SIG used for the gain measurement 
were not the same as the currently used ones. 
Especially, a SIG board with one channel was used 
when the gain was measured.
Therefore the GAIN values in the header 
represents $k \times$$g_3$, where $k$ is an unknown coefficient
of $k=g_1(old)\times$$g_2(old)$.
As the gain of the preamp used for the measurement was
$g_2(old)=4.19,$ while the typical gain of the preamps in the camera is 
$\overline{g_2}=2.57$, the factor was corrected in the FITS header values.

\begin{figure}
\includegraphics[scale=0.5,bb=0 0 378 500]{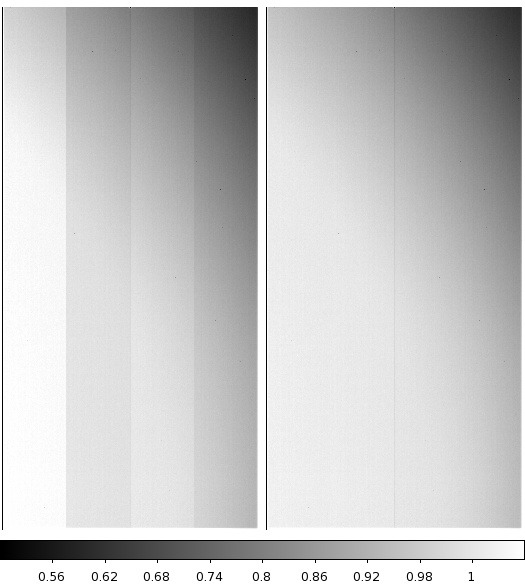}
\caption{Improvement by the new gain data. 
A V-band domeflat of the CCD0 multiplied with gain at each channel.
At left is the result with the original gain value,
and at right is that with the new values (in Table \ref{tab:gain}).
For comparison, the images are normalized so that the median of 
the image is unity, and the color scales of the two are the same.}
\label{fig:newgain}
\end{figure}

We can see the incorrect gain problem by multiplying flat images 
with the GAIN values in the FITS header at each channel. 
An example of CCD5 is shown in Fig \ref{fig:newgain} left.
The flat pattern is the product of 
the quantum efficiency (QE) of each pixel, 
the throughput pattern of optics, and the inverse of the total gain.
We can expect that the QE would be continuous at the channel boundary
within a small variation. The optics pattern should be smooth.
As there is a level gap between adjacent channels, 
it must be made by incorrect gain value ratios.
Namely, it should reflect the variation of $g_3$ among 
the channels in the same CCD.

\begin{figure}
\includegraphics[scale=0.7,bb=0 0 700 221]{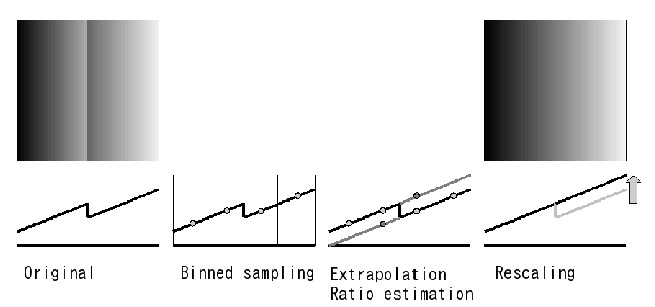}
\caption{Schematic figure of the relative gain estimation.}
\label{fig:manga_newgain}
\end{figure}

In this study, we only require the ratio of the gains,
since we expect that the crosstalk would be approximated by 
a linear relation as many other cameras follow.
We therefore recalibrate the gain of each channel using dome flat
so that the step between CCDs and channels to be minimal.
As the gain of one channel (chA) of CCD9 is changed in 2010/10,
we need to use a set before the change and after the change.
We adopted 18 exposures of V, R, and I-band taken on 2010-06-10 
for the former, and 38 exposures in V, R, and i-band 
taken between 2011-03-31 and 2011-04-04 for the latter.
Each frame is divided by the median of the frame for 
normalization. Then, the median of the normalized frames is
taken for each band and each CCD. This is a normal dome flat.
We then extracted regions of 128 pixels wide across the channel boundary 
and binned by 32$\times$32 pixels.
The left 2 binned pixels vL[1],vL[2] are in the left channel, 
and the other two pixels vR[3],vR[4] are in the right channel.
From vL[1] and vL[2], vL[3] is estimated by linear extrapolation,
and vR[2] is estimated from vR[3] and vR[4].
The ratio vR[3]/vL[3] and vR[2]/vL[2] reflect the ratio of the 
gains of the adjacent channels.
The schematic figure is in Figure \ref{fig:manga_newgain}.
As the y-pixels are 4177 in original flat, 2$\times$130 of the ratios are 
obtained. The median of the ratio gives the ratio of the gain 
of the adjacent channels in the CCD, and the error is estimated from
median of the absolute deviation (MAD).
We found that the ratio is the same within the error (typically
$\sim$0.05\%) among different bands
and in different epochs, except the chA of CCD9.
This result supports that this method well extracts 
the relative gain information, 
as the flat pattern due to optics differs in different bands.
We therefore took the median of the ratio of gains in all the bands.
Except for the chA of CCD9, the two epochs are mixed to calculate the
relative gain.

We then estimated the ratio of the gain between neighboring CCDs.
The similar method is adopted but we adopted a binning size 
of 100$\times$100 pixels, and ratios not only of x-neighbors 
but also of y-neighbors are calculated.
The relative position and rotation of CCDs were estimated 
from night sky dithered exposures. 
We adopted the positions and the rotations as in Table \ref{tab:position}.
As the gain ratio information is redundant, 
we solved the overdetermined constraints 
by a singular value decomposition method.

The relative gain values are given in Table \ref{tab:gain}.
The normalization is so that the chB of CCD5 to be unity,
for the standard stars are often observed in the channel.
The data reflect the relative values of $g_1\times$$g_2\times$$g_3$.
The application of the new values is shown in Fig \ref{fig:newgain} right
as an example, where the gaps disappear.

Dividing the re-calibrated relative gain ($g_1\times$$g_2\times$$g_3$) by
the GAIN values in the FITS header ($k \times$$g_3$), 
we can obtain relative gain of SIG+preamps, $g_1\times$$g_2$.
The values are listed in Table \ref{tab:SIGgain}.
The normalization is also at the chB of CCD5.
The change of the gain of chA of CCD9 in 2010/10 is 
thought to be the change of the on-chip amp 
and the gain of SIG+preamp remains the same.

\end{appendix}

\onecolumn
\begin{table}
\begin{tabular}{|l|c|c|c|}
\hline
EXP-ID range& number of exposures&DATE-OBS&EXPTIME[sec]\\
\hline
SUPA010806\{4,8,9\}0,SUPA01082100&4&2009-03-27&200.0\\
SUPA01085760&1&2009-03-30&120.0\\
SUPA010886[2-7]0&6&2009-04-01&300.0\\
SUPA011221[5-7]0&3&2009-08-24&180.0\\
SUPA012194[0-2]0&3&2010-04-19&240.0\\
SUPA012921[2-9]0&8&2011-03-06&300.0\\
SUPA01331150-SUPA01331260&12&2011-06-03&300.0\\
\hline
total & 37 & &\\
\hline
\end{tabular}
\caption{Used dark exposures}
\label{tab:darklist}
\end{table}

\begin{table}
\begin{tabular}{|l|c|c|c|c|}
\hline
DET-ID&chA&chB&chC&chD\\
\hline
0 & 1.050 & 1.063 & 1.083 & 1.069\\
1 & 1.070 & 1.138 & 1.217 & 1.212\\
2 & 0.995 & 1.019 & 1.038 & 0.947\\
3 & 1.022 & 1.040 & 1.042 & 1.154\\
4 & 1.006 & 1.128 & 1.085 & 1.033\\
5 & 1.015 & 1     & 0.972 & 1.082\\
6 & 1.190 & 0.987 & 0.984 & 1.034\\
7 & 0.971 & 1.059 & 1.067 & 0.981\\
8 & 1.011 & 1.252 & 1.276 & 1.172\\
9 & 0.994\tablenotemark{a},1.130\tablenotemark{b} & 1.077 & 1.039 & 1.032\\
\hline
\end{tabular}
\caption{Estimated total gain relative to chB of CCD 5}
\tablenotetext{a}{The gain is changed in 2010/10.
The data is valid for the data before the change,}
\tablenotetext{b}{The data is valid for 
the data after the change.}
\label{tab:gain}
\end{table}

\begin{table}
\begin{tabular}{|l|c|c|c|c|}
\hline
DET-ID&chA&chB&chC&chD\\
\hline
0 & 0.949 & 0.993 & 0.976 & 0.996\\
1 & 0.973 & 0.984 & 0.966 & 0.977\\
2 & 1.008 & 0.989 & 0.970 & 0.976\\
3 & 0.961 & 0.966 & 1.008 & 0.967\\
4 & 0.967 & 0.984 & 0.998 & 1.000\\
5 & 0.989 & 1     & 1.034 & 1.030\\
6 & 0.957 & 1.019 & 0.952 & 0.979\\
7 & 0.974 & 1.015 & 0.967 & 0.962\\
8 & 0.972 & 0.932 & 0.999 & 0.963\\
9 & 0.987 & 0.985 & 0.986 & 1.012\\
\hline
\end{tabular}
\caption{Estimated gain of SIG and preamp relative to chB of CCD 5}
\label{tab:SIGgain}
\end{table}

\begin{table}
\begin{tabular}{|l|c|c|c|}
\hline
DET-ID&x[pix]&y[pix]&$\theta$[rad]\\
\hline
0 & 3153.8 & 95.7 & -0.00215\\
1 & 1047.7 & 95.9 & -0.00033\\
2 & -1066.5 & 98.5 & -0.00015\\
3 & 3168.0 & -4154.4 & 0.00264\\
4 & 1050.1 & -4150.3 & -0.00020\\
5 & -1086.4 & -4158.7 & 0.00014\\
6 & -5310.5 & 96.7 & -0.00055\\
7 & -3187.3 & 97.7 & -0.00003\\
8 & -5346.6 & -4177.5 & 0.00273\\
9 & -3219.1 & -4165.6 & 0.00040\\
\hline
\end{tabular}
\caption{Relative position and rotation of CCDs}
\label{tab:position}
\end{table}

\clearpage

\begin{table}[hbt]
\begin{tabular}{|c|c|c|c|c|}
\hline
pair & N$_{\rm data}$ & $a $ & $a_{min} $ & $a_{max} $\\
\hline
gAB & 9711 & -1.42 & -1.51 & -1.33\\
gAC & 9711 & -1.48 & -1.56 & -1.40\\
gAD & 8061 & -1.67 & -1.75 & -1.59\\
gBC & 1674 & -0.62 & -0.81 & -0.44\\
\hline
\end{tabular}
\caption{
Coefficients of Best-fit regression and 
their 95\% confidence intervals (p($a_{min}<a<a_{max}$)=95\%) in $v_1$ 
in unit of $10^{-4}$}
\label{tab:coeff1}
\end{table}

\begin{table}[hbt]
\begin{tabular}{|c|c|c|c|c|}
\hline
pair & N$_{\rm data}$ & $a $ & $a_{min} $ & $a_{max}$\\
\hline
gAB & 9905 & -1.43 & -1.52 & -1.34\\
gAC & 9905 & -1.52 & -1.60 & -1.44\\
gAD & 8231 & -1.68 & -1.76 & -1.60\\
gBC & 1674 & -0.62 & -0.80 & -0.43\\
\hline
\end{tabular}
\caption{Same as Table \ref{tab:coeff1} but in $v_2$.}
\label{tab:coeff2}
\end{table}

\begin{table}[hbt]
\begin{tabular}{|c|c|c|c|c|c|c|c|c|c|c|c|c|}
\hline
& 
\multicolumn{3}{|c|}{gAB}& 
\multicolumn{3}{|c|}{gAC}& 
\multicolumn{3}{|c|}{gAD}& 
\multicolumn{3}{|c|}{gBC}\\
$\Delta$ x 
& $a$ & $a_{min}$ & $a_{max}$ 
& $a$ & $a_{min}$ & $a_{max}$ 
& $a$ & $a_{min}$ & $a_{max}$ 
& $a$ & $a_{min}$ & $a_{max}$\\
\hline
-1 & 0.12 & -0.43 & 0.67 & 0.08 & -0.42 & 0.57 & 0.01 & -0.56 & 0.59 & 0.15 & -0.65 & 0.96\\
0 & -1.48 & -1.83 & {\bf -1.13} & -1.62 & -2.00 & {\bf -1.16} & -1.67 & -2.00 & {\bf -1.28} & -0.77 & -1.57 & {\bf -0.12}\\
1 & 0.02 & -0.45 & 0.48 & 0.15 & -0.38 & 0.67 & -0.02 & -0.60 & 0.57 & 0.06 & -0.77 & 0.89\\
2 & 0.51 & {\bf 0.13} & 0.89 & 0.50 & {\bf 0.15} & 0.84 & 0.53 & {\bf 0.18} & 0.89 & 0.32 & -0.37 & 1.00\\
3 & 0.13 & -0.48 & 0.74 & 0.09 & -0.46 & 0.63 & 0.07 & -0.38 & 0.52 & 0.18 & -0.53 & 0.90\\
\hline
\end{tabular}
\caption{
Part of the values used in Figure \ref{fig:xprof_n1}
in unit of $10^{-4}$.}
\label{tab:xprof_n1}
\tablenotetext{*}{The values significantly different from 0,
i.e. $a_{max}<0$ and $a_{min}>0$, are in bold font.}
\end{table}

\clearpage

\clearpage

\begin{figure}
\includegraphics[scale=0.5,bb=0 0 351 190]{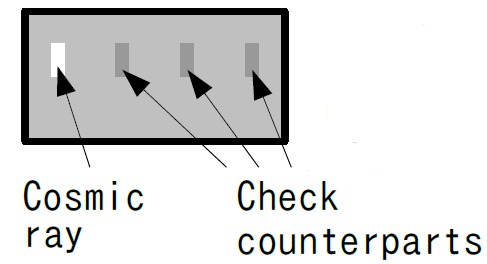}
\caption{A schematic figure of the idea of our method
to estimate the crosstalk effect.
}
\label{fig:manga_method}
\end{figure}

\clearpage

\clearpage

\clearpage

\clearpage

\clearpage

\clearpage

\clearpage

\clearpage

\end{document}